\documentclass[twocolumn,aps,floatfix,superscriptaddress,prl,showpacs]{revtex4}
\input epsf.sty
\usepackage{graphicx}
\usepackage{dcolumn}
\usepackage{bm}

\begin{document}
\title{Surface criticality at a dynamic phase transition}
\author{Hyunhang Park and Michel Pleimling}
\affiliation{Department of Physics, Virginia Tech, Blacksburg, VA 24060-0435, USA}

\begin{abstract}
In order to elucidate the role of surfaces at nonequilibrium phase transitions we consider
kinetic Ising models with surfaces subjected to a periodic oscillating magnetic field. Whereas
the corresponding bulk system undergoes a continuous nonequilibrium
phase transition characterized by the exponents of the equilibrium Ising model, we find
that the nonequilibrium surface exponents do not coincide with those of the equilibrium
critical surface. In addition, in three space dimensions the surface phase diagram
of the nonequilibrium system differs markedly from that of the equilibrium system.
\end{abstract}


\maketitle

The ubiquity of nonequilibrium steady states in nature constitutes a permanent reminder
of the challenges encountered when trying to understand interacting many-body systems far
from equilibrium. Whereas in some instances, as for example paradigmatic transport models \cite{Cho11}
or driven diffusive systems \cite{Sch95}, notable progress has been achieved in understanding
nonequilibrium steady states, a common theoretical framework remains elusive. This
is especially true in cases where steady states are influenced by the presence of
surfaces or interfaces, which can change properties even deep inside the bulk
\cite{Jan88, Har88, Ess96,Fro01, Ple04a, Ple04b, Ple05, Bau07, Kad08,Huc09, Igl11, Agl07, Agl11,Ple10,Li12}.

Nonequilibrium phase transitions form an interesting class of phenomena that share many
commonalities with their equilibrium counterparts. For example, for continuous transitions
different universality classes, characterized by different sets of critical exponents, have been
identified. Well-known examples can be found in driven diffusive systems \cite{Sch95},
at absorbing phase transitions \cite{Hen08,Odo08} or in magnetic systems subjected to a periodically oscillating
external field \cite{Cha99,Ach05}. For some absorbing phase transitions, as for example directed
percolation, the surface critical properties have been studied to some extent, see \cite{Fro01}
and references therein.

Kinetic ferromagnets in a periodically oscillating magnetic field display as a function of the
field frequency a nonequilibrium phase transition between a dynamically disordered phase
at low frequencies and a dynamically ordered phase at high frequencies. Let us assume that the magnetization
is aligned with the direction of the external field. If the field now reverses direction, the system
becomes metastable and tries to reverse its magnetization through the nucleation of droplets that are
aligned with the field. If the period of the field is large compared to the metastable lifetime, then
the metastable state completely decays before the field reverses direction again, i.e. the ferromagnet is able
to 'follow' the field, yielding a time-dependent magnetization that oscillates symmetrically about zero.
The dynamically ordered phase is obtained when the period of the field is small compared to the metastable lifetime,
thus that the system is not able to fully decay from the metastable state before the field changes direction
again. The magnetization then oscillates about a non-zero value. This behavior has been studied theoretically
in a large range of systems, as for example the Ising \cite{Ach97,Sid98,Kor00}, Heisenberg \cite{Jan03,
Ach03}, or Blume-Emery-Griffiths \cite{Can06} models, too name but a few.
Possible experimental realizations have been discussed in Co films on Cu(001) \cite{Jia95} as well as in 
[Co/Pt]$_3$ magnetic multilayers \cite{Rob08}. Of special interest in the following is the kinetic Ising model
in an oscillating field that is displaying critical exponents at a
dynamic phase transition which are identical to those found at the phase transition of the
equilibrium Ising model \cite{Kor00}. This surprising observation is consistent with a
symmetry argument given in \cite{Gri85} and has been substantiated through the study of the time-dependent
Ginzburg-Landau model in an oscillating field \cite{Fuj01}.

In the past very few studies have looked at the impact surfaces can have on this dynamic phase transition, 
thereby focusing mostly on rather general aspects. For example, the effects of boundaries on
magnetization switching in kinetic Ising models were studied in \cite{Ric97}.
In \cite{Jan01,Jan03} the dynamic phase transition was investigated in Heisenberg films with 
competing surface fields.

In this Letter we present the first study of the surface critical properties at a dynamic 
phase transition. Using large-scale numerical simulations, we study kinetic
Ising models with free surfaces subjected to a square-wave oscillating field.
Both in two and three space dimensions we obtain values for the surface critical exponents that differ
markedly from the values of the equilibrium surface exponents, thus demonstrating that the dynamic
surface universality class differs from that of the equilibrium system, even though the same universality
class prevails for the corresponding bulk systems. In addition, we find that the kinetic surface phase diagram
in three dimensions is remarkably simple and does not exhibit a special transition point, nor a surface or
extraordinary transition, which are all present in the equilibrium surface phase diagram.

In order to study the surface critical behavior at the dynamic phase transition
we consider square and cubic lattices
with open boundary conditions in one direction, called $z$-direction
in the following, whereas in the direction(s) perpendicular
to the $z$-direction we have periodic boundary conditions \cite{Ple04}. In this way we have in a system 
of linear extend $L$ two
surfaces, located at $z =1$ and $z=L$. Every lattice site $\textbf{x}$ is characterized by
an Ising spin $S_{\textbf{x}} = \pm 1$. The Hamiltonian is given by
\begin{equation}
H=-J_b\sum_{\langle
\textbf{x},\textbf{y}\rangle}S_{\textbf{x}}S_{\textbf{y}}-J_s\sum_{\lbrace
\textbf{x},\textbf{y}\rbrace}S_{\textbf{x}}S_{\textbf{y}}-H(t)\sum_{
\textbf{x}}S_{\textbf{x}}~,
\end{equation}
where $J_b > 0$ and $J_s >0$ are ferromagnetic bulk and surface coupling constants. 
The first sum is over nearest neighbor
sites where at most one of the sites is in a surface layer. The second sum, on the other hand, is 
over neighboring sites that are both in a surface layer. We thereby allow for different 
values of the coupling constants
at the surface and inside the bulk. Finally, the last term is the magnetic field term where
$H(t)$ is a spatially uniform field that oscillates in time. We follow \cite{Kor00} and
use a square-wave field with amplitude $H_0$. Both temperature and magnetic field strength are
chosen in such a way that the system is in the multidroplet regime \cite{Kor00}: 
$T=0.8T_c^{2d}$, $H_0=0.3J_b$ for $d=2$ and $T=0.8T_c^{3d}$, $H_0=0.4J_b$ for $d=3$. 
Here $T_c^{2d} = 2.269 \cdots J_b/k_B$
and $T_c^{3d} = 4.5115 J_b/k_B$ are the critical temperatures of the two- and
three-dimensional equilibrium systems.

As the surfaces break spatial translation invariance, all quantities of interest
depend on the distance to the surface. We therefore define local, i.e. layer-dependent,
quantities. Thus we consider the layer
magnetization averaged over one period of the external field ($t_{1/2}$ is the half-period
of the oscillating field):
\begin{equation} \label{eq:Q}
Q(z)=\frac{1}{2t_{1/2}}\oint m(t,z)dt~,
\end{equation}
with the time-dependent magnetization $m(t,z)=\frac{1}{L^{d-1}}\sum_{\textbf{x}}S_{\textbf{x}}(t)$
of layer $z$, the sum being taken over all spins in that $(d-1)$-dimensional layer. 
The local order parameter is then given by $\langle | Q(z) | \rangle$ where $\langle \cdots \rangle$
indicates both a time average (i.e. an average over many periods) and a thermal average
(realized in the numerical simulations through multiple independent runs with different
random number sequences), yielding typically a total of 500,000 periods over which the average is taken. 
In a similar way we define the layer Binder cumulant
\begin{equation} 
U(z)=1-\frac{\langle \left( Q(z) \right)^4\rangle}{3\langle \left( Q(z) \right)^2\rangle^2}
\end{equation}
and the layer susceptibility
\begin{equation} 
\chi(z)=L^{d-1}\left(\langle Q(z)^2\rangle-\langle|Q(z)|\rangle^2\right)~.
\end{equation}
In the following the surface quantities will be characterized by an index {\it s}, whereas an
index {\it b} will be given to the quantities from the middle of the sample.

An important quantity in the study of the dynamic phase transition is the ratio
\begin{equation} 
\Theta=\frac{t_{1/2}}{\langle\tau\rangle_b}
\end{equation}
that quantifies the competition between the metastable state, characterized by the metastable
lifetime $\langle\tau\rangle_b$, and the oscillating magnetic field. For small values of $\Theta$
we are in the dynamically ordered phase, whereas for large values the system is
dynamically disordered. The quantity $\Theta$ therefore plays the same role as that
played by temperature at an ordinary equilibrium phase transition. The metastable lifetime in
our systems is again layer dependent, the value of $\langle\tau(z)\rangle$ being smaller
in the surface layer than deep inside
the bulk, as surface spins are coupled to fewer spins. As we are
interested in the surface properties at the bulk phase transition, we define $\Theta$ with
respect to the bulk quantity $\langle\tau\rangle_b$. The dynamic phase transition then takes
place at the critical value $\Theta_c = 0.918$ in $d=2$ \cite{Kor00} and $\Theta_c = 1.285$
in $d=3$.

As shown in Figure 1 for the two-dimensional system with $J_s = J_b$, both the bulk and the surface 
order parameters decrease rapidly when approaching the critical point $\Theta_c$ from below.
Concomitantly, the bulk and surface susceptibilities display peaks in the vicinity of $\Theta_c$.
Changing the system size yields system size dependencies (shifts of the positions of the maxima of
the susceptibilities, increasing peak heights, $\cdots$) typical for a continuous 
phase transition.

%%%%%%%%%%%%%%%%%%%%%%%%%%%%%%%%%%%%%%%%%%%FIG 1.%%%%%%%%%%%%%%%%%%%%%%%%%%%%%%%%%%%%%%%%%%%%%%%%%%%%%%
\begin{figure} [h]
\includegraphics[width=0.90\columnwidth]{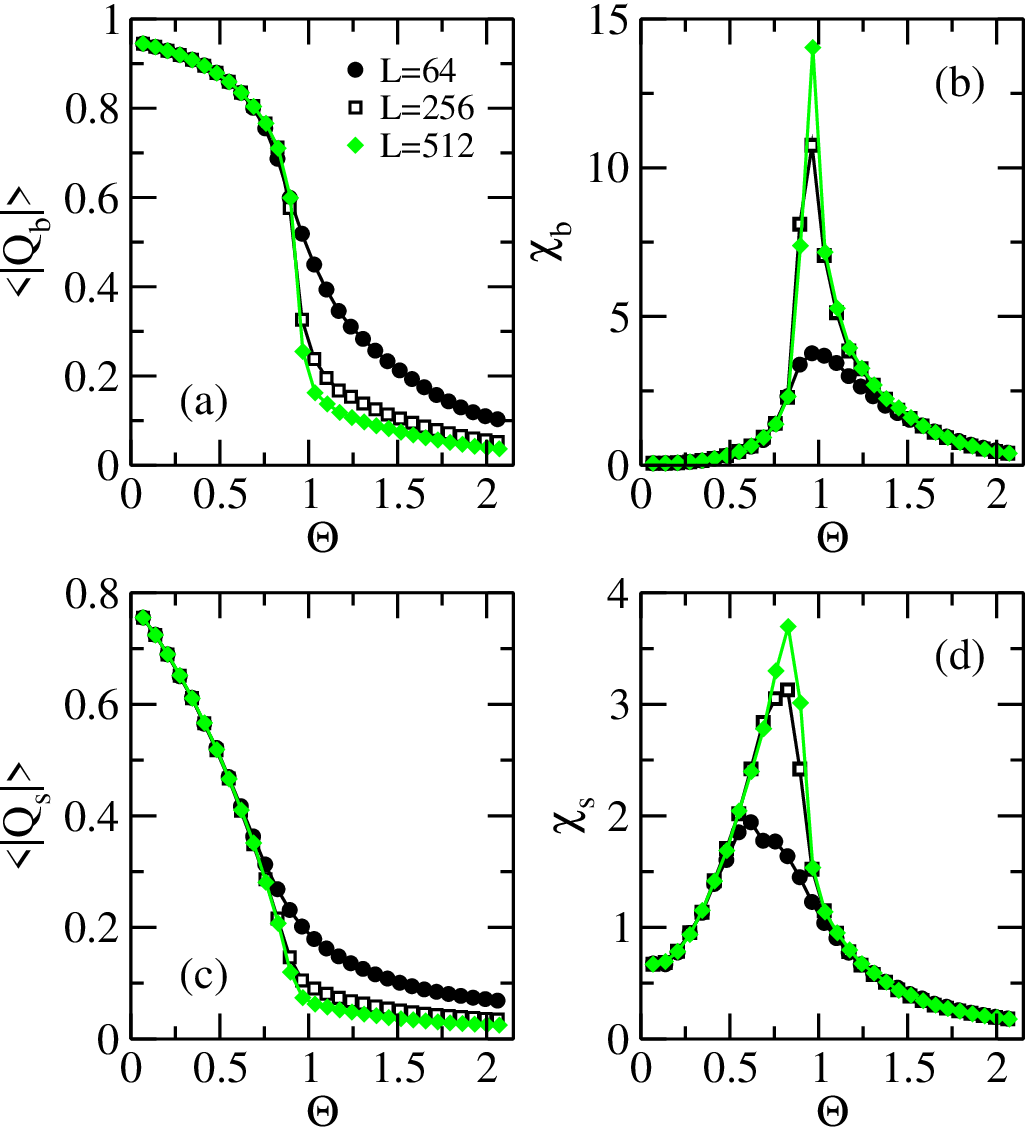}
\caption{\label{fig1} (Color online) 
Bulk (a,b) and surface (c,d) quantities for the two-dimensional model, composed
of $L \times L$ spins, with $J_s = J_b$. Close to the bulk critical point $\Theta_c = 0.918$,
the local order parameters decrease rapidly and 
the local susceptibilities display pronounced maxima. Here and in the following error bars are
smaller than the sizes of the symbols.
}
\end{figure}
%%%%%%%%%%%%%%%%%%%%%%%%%%%%%%%%%%%%%%%%%%%FIG 1.%%%%%%%%%%%%%%%%%%%%%%%%%%%%%%%%%%%%%%%%%%%%%%%%%%%%%%

In the infinite system, the order parameter and the susceptibility show an algebraic behavior
close to the critical point:
\begin{equation}
\langle | Q_b | \rangle \sim (\Theta_c - \Theta )^\beta ~~,~~ \chi_b \sim \left| \Theta_c - \Theta
\right|^{-\gamma}~.
\end{equation}
Using finite-size scaling \cite{Bin90}, the authors of \cite{Kor00} found in
two dimensions the same values for the exponents as those obtained for the equilibrium
Ising model, namely $\beta/\nu = 1/8$ and $\gamma/\nu = 7/4$, where $\nu$ is the critical exponent
that governs the divergence of the correlation length.

\begin{figure} [h]
\includegraphics[width=0.90\columnwidth]{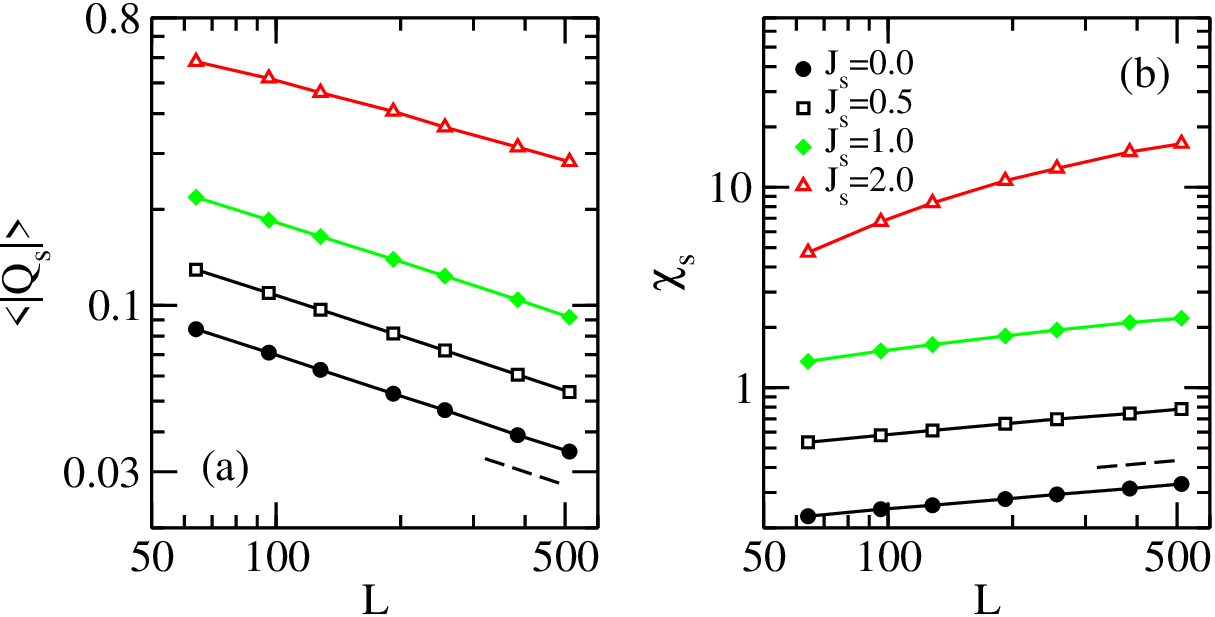}
\caption{\label{fig2} (Color online)
Log-log plot of (a) the surface order parameter and (b) the surface susceptibility 
as a function of the linear system size
for the two-dimensional kinetic Ising model at $\Theta = \Theta_c$.
The different curves correspond to different values of the surface coupling constant $J_s$ (given
in units of $J_b$). The dashed lines have slopes $-0.43$ (a) and 0.18 (b). For large values of $J_s$
corrections to scaling become sizeable.
}
\end{figure}

Similarly, surface critical exponents are introduced to describe the behavior of surface quantities
close to the bulk critical point (we use here the standard nomenclature of surface critical
phenomena, see \cite{Ple04,Bin83,Die86}):
\begin{equation}
\langle | Q_s | \rangle \sim (\Theta_c - \Theta )^{\beta_1} ~~,~~ \chi_s \sim \left| \Theta_c - \Theta
\right|^{-\gamma_{11}}~.
\end{equation}
Close to a bulk critical point, finite-size scaling theory \cite{Ple04,Bin83,Die86} provides us
with scaling relations for our surface quantities:
\begin{eqnarray}
\langle|Q_s|\rangle&=&L^{-\beta_1/\nu}F_{\pm}(\theta L^{1/\nu})\\
\chi^Q_s&=&L^{\gamma_{11}/\nu}G_{\pm}(\theta L^{1/\nu})
\end{eqnarray}
where $\theta=\frac{|\Theta-\Theta_c|}{\Theta_c}$, whereas $F_{\pm}$ and $G_{\pm}$ are scaling functions,
where the $+$ ($-$) sign corresponds to $\Theta >$ ($<$) $\Theta_c$. Choosing $\Theta=\Theta_c$, we therefore
expect that our quantities depend algebraically on the linear system size. This is shown in Figure
\ref{fig2} for various values of the surface coupling $J_s$. For not too large values of $J_s$ 
corrections to scaling are negligible, so that we can determine the values of the critical exponents
from the slopes. We find $\beta_1/\nu = 0.43(1)$ and $\gamma_{11}/\nu = 0.18(1)$.
We immediately remark that these values rather well fulfill the scaling relation $2 \beta_1 + \gamma_{11} = d-1$ that
is expected to hold for surface critical exponents. We also note that our values differ strongly 
from the values of the surface exponents in the equilibrium critical Ising model:
$\beta_1^e/\nu = 0.5$ and $\gamma_{11}^e/\nu = 0$ \cite{Ple04,Bin83,Die86}. 
We therefore have the interesting situation that
while the dynamic phase transition in the bulk belongs to the universality of the equilibrium Ising model,
this is not true for the corresponding surface universality class.

%%%%%%%%%%%%%%%%%%%%%%%%%%%%%%%%%%%%%%%%%%%FIG 2.%%%%%%%%%%%%%%%%%%%%%%%%%%%%%%%%%%%%%%%%%%%%%%%%%%%%%%
%%%%%%%%%%%%%%%%%%%%%%%%%%%%%%%%%%%%%%%%%%%FIG 2.%%%%%%%%%%%%%%%%%%%%%%%%%%%%%%%%%%%%%%%%%%%%%%%%%%%%%%

\begin{figure} [h]
\includegraphics[width=0.90\columnwidth]{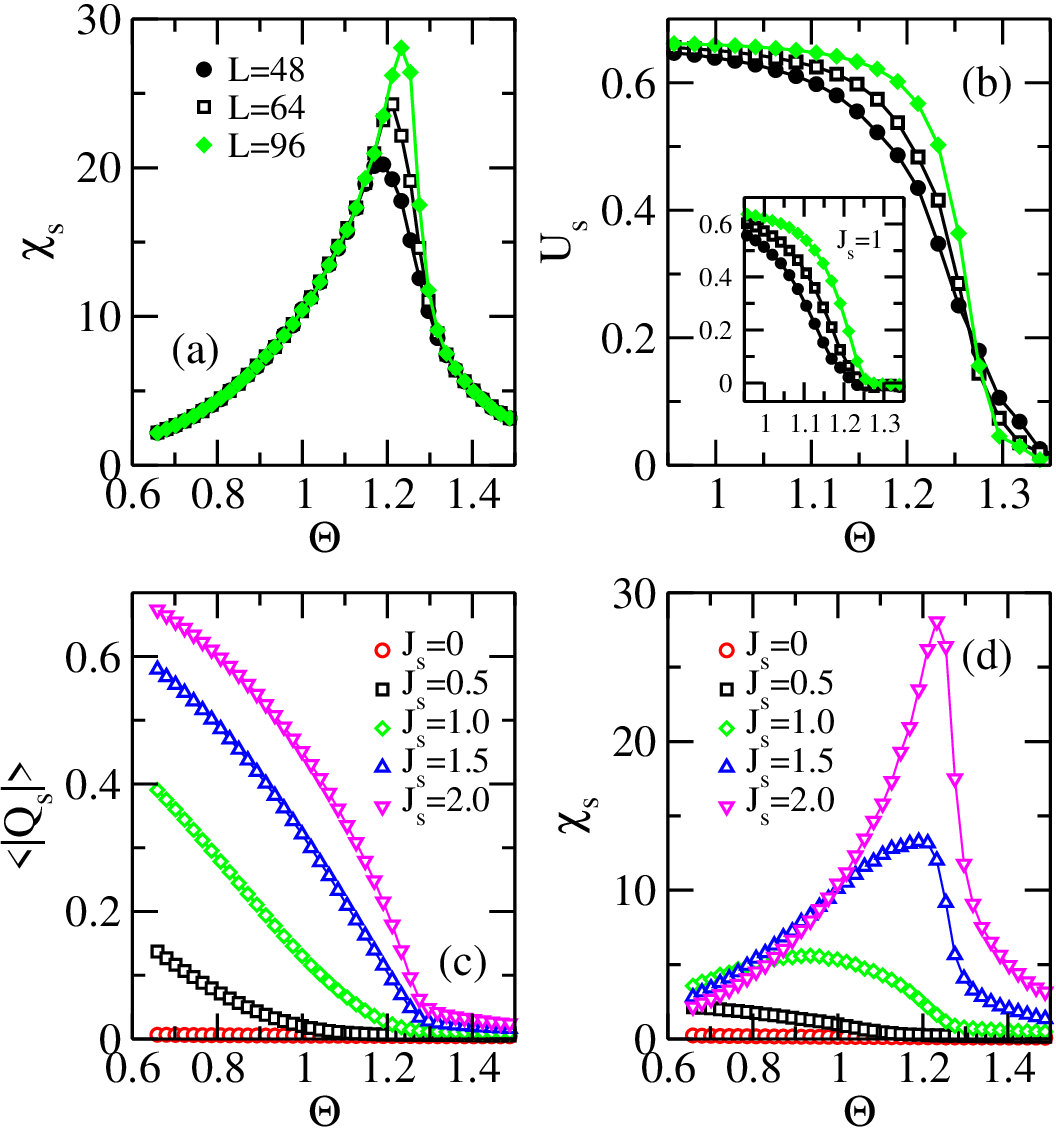}
\caption{\label{fig3} (Color online)
Surface susceptibility (a) and surface Binder cumulant (b) for the three-dimensional kinetic Ising
model with $J_s = 2 J_b$. For small values of $J_s$, the surface does not order dynamically at the
bulk critical point, as shown by (c) the surface order parameter and (d) the surface susceptibility.
All surface coupling constants are expressed in units of $J_b$. The inset in (b) shows that for
$J_s = J_b$ the surface Binder cumulants do not cross at a common value of $\Theta$. The system size
in (c) and (d) is $L =96$.
}
\end{figure}

For the three-dimensional system, the bulk system undergoes again a dynamic
phase transition characterized by the critical exponents of the three-dimensional equilibrium
critical Ising model, as we verified. For the surface, however, the situation is more complicated. 
For not too small values of the
surface coupling, the situation is similar to the two-dimensional case, see the example
$J_s = 2 J_b$ shown in the first row in Fig. \ref{fig3}:
When reducing $\Theta$ the surface undergoes at the bulk transition value $\Theta_c$ a transition to a dynamically 
ordered phase. This transition is revealed by a characteristic peak in the surface susceptibility as well
as by a crossing at $\Theta_c$ of the surface Binder cumulant computed for different system sizes.
However, for values of $J_s < 1.5 J_b$, see Fig.\ \ref{fig3}c, 
the surface spins do {\it not} order dynamically at $\Theta_c$, 
but instead are still able of following the external field, even though this is no longer the case for
the bulk spins. At lower values of $\Theta$ the surface order parameter of our finite systems deviates
from zero, but this partial dynamical ordering is not related to a phase transition. This is also
revealed by the presence of a non-critical peak (or, for very small values of $J_s$, by the complete absence
of any peak) in the surface susceptibility (see Fig. \ref{fig3}d) as well as by the absence of the crossing
of the surface Binder cumulants for different system sizes, as shown in the inset of Fig. \ref{fig3}b.

%%%%%%%%%%%%%%%%%%%%%%%%%%%%%%%%%%%%%%%%%%%FIG 3.%%%%%%%%%%%%%%%%%%%%%%%%%%%%%%%%%%%%%%%%%%%%%%%%%%%%%%
%%%%%%%%%%%%%%%%%%%%%%%%%%%%%%%%%%%%%%%%%%%FIG 3.%%%%%%%%%%%%%%%%%%%%%%%%%%%%%%%%%%%%%%%%%%%%%%%%%%%%%%

\begin{figure} [h]
\includegraphics[width=0.90\columnwidth]{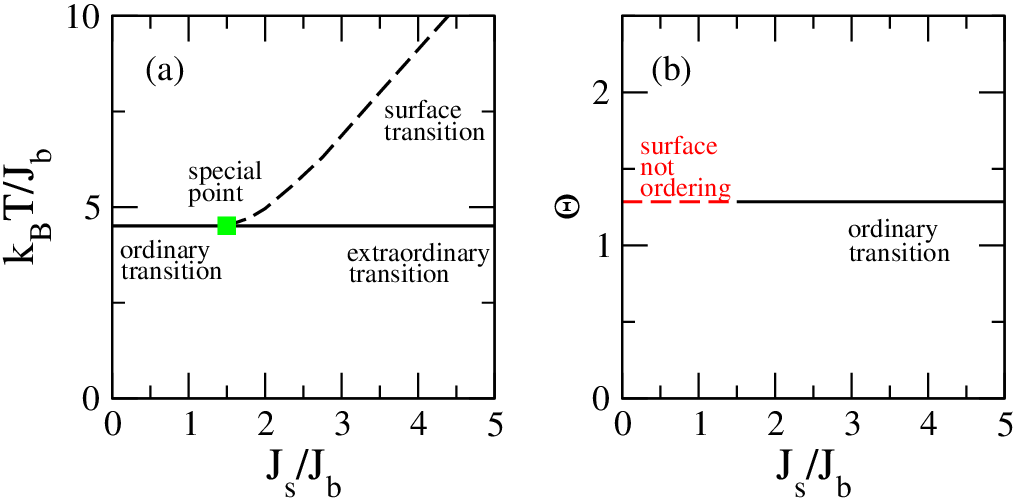}
\caption{\label{fig4} (Color online)
Surface phase diagram of (a) the equilibrium three-dimensional Ising model and (b) the nonequilibrium
three-dimensional kinetic Ising model.
}
\end{figure}

Based on our data, where we studied surface couplings from $J_s =0$ to $J_s = 16 J_b$,
the surface phase diagram of the three-dimensional kinetic Ising model 
shown in Fig. \ref{fig4}b differs remarkably from the corresponding diagram of the equilibrium
model \cite{Ple04,Bin83,Die86}, see Fig. \ref{fig4}a. Not only does the surface not order dynamically
at the bulk critical point for $J_s < 1.5 J_b$, as just discussed, the kinetic Ising model also
does not exhibit a surface transition, where the surface orders alone, whereas the bulk remains
disorder. Concomitantly, the special transition point, where both surface and bulk are critical, 
and the extraordinary transition, where the bulk orders in presence of an ordered surface, are also
absent. In fact, whereas for the equilibrium system is it possible to shift the phase transition temperature
of the two-dimensional surface $k_B T_c^{2d} = 2.269 \cdots J_s$ above the bulk transition temperature
$k_B T_c^{3d} = 4.5115 J_b$ by sufficiently increasing the ratio $J_s/J_b$ of the couplings, a similar
mechanism does not exist in the kinetic Ising model. 

%%%%%%%%%%%%%%%%%%%%%%%%%%%%%%%%%%%%%%%%%%%FIG 4.%%%%%%%%%%%%%%%%%%%%%%%%%%%%%%%%%%%%%%%%%%%%%%%%%%%%%%
%%%%%%%%%%%%%%%%%%%%%%%%%%%%%%%%%%%%%%%%%%%FIG 4.%%%%%%%%%%%%%%%%%%%%%%%%%%%%%%%%%%%%%%%%%%%%%%%%%%%%%%

\begin{figure} [h]
\includegraphics[width=0.90\columnwidth]{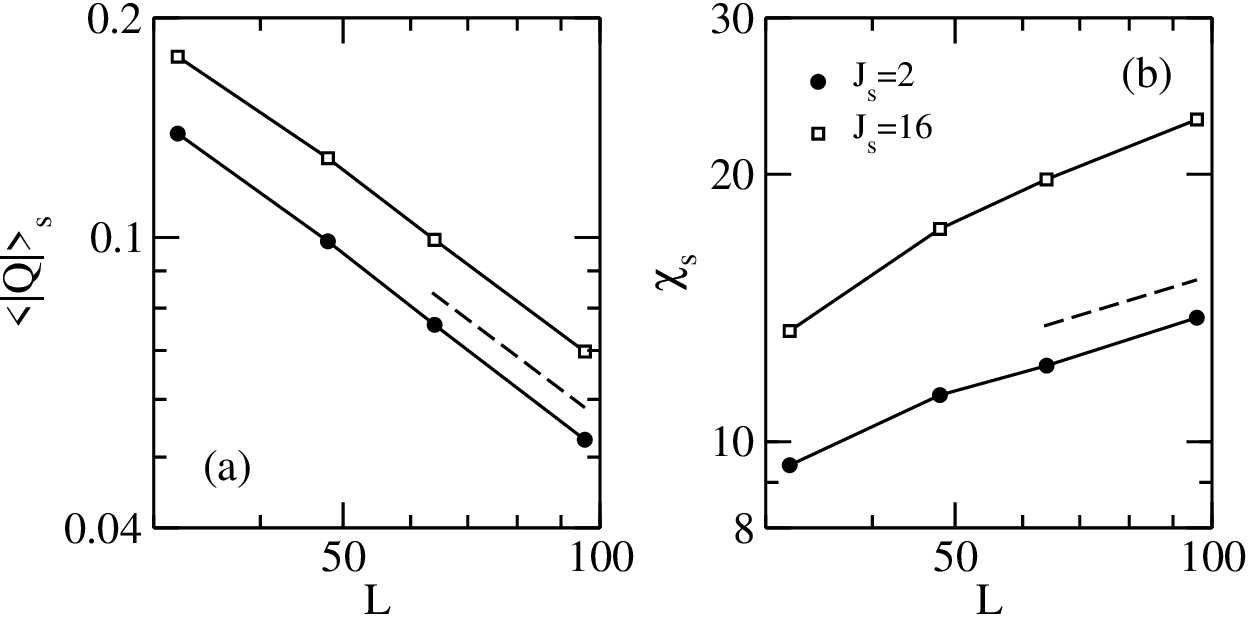}
\caption{\label{fig5} 
(a) Surface order parameter and (b) surface susceptibility as a function of system size at the
bulk critical point of the three-dimensional kinetic Ising model. 
The dashed lines have slopes $-0.88$ (a) and 0.29 (b).
}
\end{figure}

Finally, for $J_s > 1.5 J_b$ we can again measure the surface critical exponents through
a finite-size scaling analysis. As shown in Fig. \ref{fig5}
corrections to scaling are much more important in three than in two dimensions. Based on our data,
we obtain $\beta_1/\nu = 0.88(3)$ and $\gamma_{11}/\nu = 0.29(3)$. These values again differ markedly
from the known values $\beta^e_1/\nu = 1.27$ and $\gamma^e_{11} = -0.40$ 
of the corresponding surface critical exponents \cite{Die98}. Most notably, whereas in the equilibrium system
the surface susceptibility displays a cusp singularity characterized by a negative critical
exponent, in our system the surface susceptibility diverges with a positive critical exponent.

%%%%%%%%%%%%%%%%%%%%%%%%%%%%%%%%%%%%%%%%%%%FIG 5.%%%%%%%%%%%%%%%%%%%%%%%%%%%%%%%%%%%%%%%%%%%%%%%%%%%%%%
%%%%%%%%%%%%%%%%%%%%%%%%%%%%%%%%%%%%%%%%%%%FIG 5.%%%%%%%%%%%%%%%%%%%%%%%%%%%%%%%%%%%%%%%%%%%%%%%%%%%%%%

As we mentioned in the introductory remarks, a symmetry argument put forward in \cite{Gri85} states
that continuous transitions with up-down symmetry and nonconserved order parameter should fall into
the university class of the ordinary Ising model. This indeed agrees with our own results (as well as
with previous results \cite{Kor00,Fuj01}) that the bulk critical exponents at the dynamic phase transition
are the same as that of the equilibrium Ising model, and this both in two and three dimensions. 
However, once surfaces are introduced, the lattice symmetry is broken close to the surfaces, and one
of the assumptions underlying the argument of \cite{Gri85} is no longer fulfilled. Indeed, our results
show that the dynamic surface exponents differ from the surface exponents of the equilibrium model,
yielding new nonequilibrium surface universality classes. Using field-theoretical methods similar to those
developed for equilibrium critical surfaces \cite{Die86,Die98}, it should be possible to compute these new exponents 
and to classify the possible dynamic surface universality classes.

It follows from our work that our understanding of the role played by surfaces in nonequilibrium systems,
and more specifically at nonequilibrium phase transitions, is far from being complete. Surfaces break
lattice symmetries, and this can have many surprising and unexpected effects out of equilibrium, as 
exemplified in our study of surface critical behavior at a dynamic phase transition. Based on our
results we expect that future in-depth 
studies of the role of surfaces far from equilibrium will reveal additional new and unexpected
phenomena.

This work was supported by the US National
Science Foundation through DMR-0904999.

\end{document}